**Bracing for sustainable agriculture: the development and function of brace roots in members of *Poaceae***

Ashley N. Hostetler[1]*, Rajdeep S. Khangura[2]*, Brian P. Dilkes[2], and Erin E. Sparks[1]

**Affiliations**:
[1]Department of Plant and Soil Sciences and the Delaware Biotechnology Institute, University of Delaware, Newark, DE 19711
[2]Department of Biochemistry, Purdue University, West Lafayette, IN 47907
*equal contribution

**Corresponding Author** esparks@udel.edu

**Contributions** All authors contributed to the writing and editing of the manuscript

# Highlights

- Aerial nodal roots (brace roots) are important in plant anchorage and consequently lodging resistance
- In a meta-analysis, more than half of the genes known to influence brace roots in maize are also characterized to alter the juvenile-to-adult phase change
- Agriculturally important C4 crops such as maize, sorghum, sugarcane, and *Setaria* develop aerial nodal roots, however, studies of the development and function remain limited

# Graphical Abstract

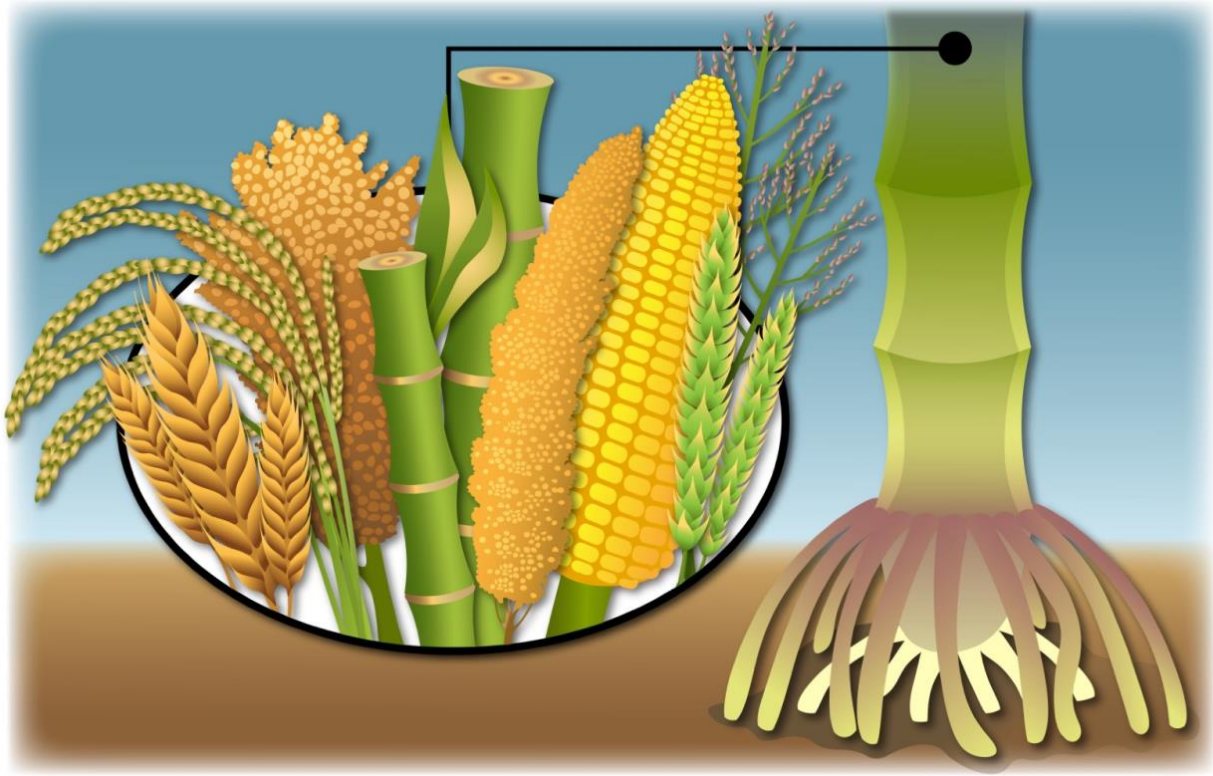

# Abstract

Optimization of crop production requires root systems to function in water uptake, nutrient use, and anchorage. In maize, two types of nodal roots–subterranean crown and aerial brace roots function in anchorage and water uptake and preferentially express multiple water and nutrient transporters. Brace root development shares genetic control with juvenile-to-adult phase change and flowering time. We present a comprehensive list of the genes known to alter brace roots and explore these as candidates for QTL studies in maize and sorghum. Brace root development and function may be conserved in other members of *Poaceae*, however research is limited. This work highlights the critical knowledge gap of aerial nodal root development and function and suggests new focus areas for breeding resilient crops.

# Introduction

Plant adaptation to changing environments requires the dynamic response of root systems to optimize water uptake, nutrient use, and anchorage [1–4]. Recent work highlights the importance of aerial roots in plant adaptation. Exploring the development and function of aerial roots can fill a critical knowledge gap and identify new targets for breeding resilient crops through root system optimization.

The maize (*Zea mays*) aerial root system presents an ideal target for this research. The maize root system consists of different root types including embryonic primary and seminal roots that are subterranean, and postembryonic nodal roots that are both subterranean and aerial (**Figure 1**)[4–8]. Nodal roots, a type of adventitious root, develop sequentially from stem nodes rather than the embryo or other roots (i.e. lateral roots). Two classifications of nodal roots exist in maize – those initiating from subterranean stem nodes, referred to as crown roots, and those initiating from aerial stem nodes, referred to as brace roots (**Figure 1**)[4,6–8]. Typically, five stem nodes belowground create a dense rootstock of crown roots, whereas the brace roots initiate from two or three stem nodes above the soil line [9]. The number of nodes with emerged brace roots is influenced by both genotype and environment [10–12].

From a functional perspective, embryonic roots in maize provide nutritional and structural support early in seedling establishment. However, after maize plants reach the vegetative leaf 6 (V6) stage, nutritional and structural support is primarily provided by the subterranean and aerial nodal roots [4]. Aerial nodal roots were historically observed to prop (or 'brace') the plant in the event of lodging (crop failure). Although many nodes above the soil produce brace roots in an acropetal sequence only the first few nodes have roots that enter the soil, a trait that has recently been shown to be important for anchorage [12–14]. Optimizing the function of these roots for sustainable agriculture requires a detailed understanding of the developmental mechanisms that regulate their formation.

# Nodal Root Development

Nodal roots are adventitious roots, which develop in four sequential stages: induction, initiation, emergence, and growth (**Figure 1**)[8,15]. Induction requires one or more cell types to become founder cells with the competency to divide (**Figure 1A**). This stage is poorly understood for the development of adventitious roots and open questions remain as to which cell(s) or tissue(s) initiate adventitious roots and if founder cell establishment requires dedifferentiation [8,15]. Typically, once founder cells are established, they are competent to begin initiation [15]. The initiation stage is when cell division leads to observable morphological changes (i.e. presence of primordia). At the end of initiation, mature root primordia are established within the stem node (**Figure 1B**). The primordia remain in this state until further signaling promotes emergence. In the emergence stage, root primordia elongate and emerge from the node (**Figure 1C**). After emergence, root growth determines the extent of the root system, including the number of nodes containing brace roots that enter the soil (**Figure 1D**). Development can be regulated at each of these stages, which may have divergent regulation based on the plant species and adventitious root type [16].

The developmental mechanisms that regulate the first two stages of development (induction and initiation) are likely shared between subterranean crown roots and aerial brace roots. Both roots are initiating from stem nodes, which are anatomically comparable. However, the regulation of emergence and growth is likely influenced by distinct environmental cues based on the subterranean or aerial environment. This divergence is supported by different anatomy for the portion of a brace root that is in the air versus the portion that has entered soil [9]. The literature often reports collective nodal root data without distinguishing subterranean or aerial roots, which has limited the discovery of distinct developmental mechanisms.

In maize, nodal root development has been characterized through screening mutants for root defects. These screens have identified mutants affecting different developmental stages of adventitious root development. Loss-of-function mutants in two paralogous genes of maize, *rootless concerning crown and seminal roots1 (rtcs1)* and *rtcs-like1 (rtcl1)*, exhibit root development defects. Loss of the more abundantly expressed, *rtcs1* inhibits seminal, crown, and brace root primordia formation, indicating a defect in founder cell establishment or cell division (**Figure 1, Table 1**)[17]. A mutant of *rtcs-like1* (*rtcl1*) has a weaker phenotype that reduces nodal root growth after emergence (**Figure 1D**) [18]. Loss of the rice ortholog of *rtcs1/rtcl1*, *Crown rootless1* (*Crl1/Arl1*), blocks nodal root initiation [19–21] demonstrating a shared mechanism for nodal root induction or initiation. Another recessive mutant of maize, *lateral rootless1* (*lrt1*), is defective in the formation of primordia at the first coleoptile node suggesting divergent induction or initiation mechanisms at different nodes. The *lrt1* mutants form other nodal roots but these roots have reduced growth after emergence (**Figure 1D**) [22]. Additional signals are required to promote emergence. For example, overexpression of *auxin regulated gene involved in organ size8* (*argos8*) in maize blocks nodal root emergence, but does not impact primordia formation (**Figure 1C**) [13]. While other mutants that lack nodal roots have been identified, the stage of development that is inhibited has yet to be defined. Specifically, the basic helix-loop-helix (bHLH) transcription factor mutant *rootless1* (*rt1*) [23,24] and the AP2/ERF transcription factor mutant *zmrap2.7* [25] have a reduced number of nodes with roots, whereas the MATE transporter mutant *big embryo1* (*bige1*) [26] has an increased number of nodes with roots (**Supplemental Table 1**).

These genes encompass hormone pathways and transcriptional regulators. For example, the expression of RTCS1 is induced by auxin [27], a plant hormone which is essential for initiation of other stem-born roots [28–37]. The absence of primordia in the *rtcs1* mutant indicates that auxin is also a key regulator of nodal root induction or initiation in maize. In contrast, the accumulation of ARGOS8 inhibits ethylene signaling and reduces nodal root emergence (**Figure 1C**) [13]. These data suggest that auxin and ethylene have distinct functions during maize nodal root development. While the interplay between auxin and ethylene in adventitious root development has been previously defined [16], the exact mechanism is highly context dependent. Several transcriptional regulators have been defined by mutant analyses, however there is very little known about the molecular mechanisms of these genes in relation to nodal root development. Maize nodal roots emerge from juvenile nodes [38]. Thus, one approach to uncover the molecular mechanisms of brace root development is to consider this link to juvenility.

## Phase Change, Flowering Time, and Brace Roots in Maize

The juvenile-to-adult and adult-to-reproductive transitions bring major morphological and physiological changes in plant development [39,40]. The juvenile-to-adult transition refers to two phases of vegetative development. The transition to the adult vegetative phase is characterized by changes in morphology and the acquisition of competence of the shoot meristem to respond to reproductive signals. The subsequent adult-to-reproductive transition is the initiation of the reproductive meristem development and the transition from vegetative to reproductive organ initiation [39]. Although these two transition phases are often separated by time in perennial plants, in annual plants (e.g. maize) adult vegetative transition and reproductive transition are sometimes confused because meristem competence can occur simultaneously in the presence of reproductive signals [41].

The juvenile-to-adult phase change in flowering plants is well conserved. This conservation is highlighted by regulation of gene expression through two microRNAs (miR), miR156 and miR172, and their respective mRNA targets, SQUAMOSA PROMOTER BINDING PROTEIN-LIKE (SPL) and APETALA2-LIKE MADS (AP2-LIKE) transcription factors [42]. miR156 represses SPL genes to retain juvenile identity in Arabidopsis, maize, and multiple woody perennial species [40,43,44]. After the transition to the adult phase, SPL transcription factors promote miR172 expression, which represses AP2-LIKE transcription factors to initiate the reproductive phase [45–48]. The emergence of brace roots from juvenile nodes may be attributed to the direct function of the miR156/SPL signaling module or indirectly as an outcome of node identity. This signaling module directly controls lateral root development in Arabidopsis independent of phase change [49]. However, repression of miR156 in Arabidopsis hypocotyls reduced the number of adventitious roots, suggesting that adventitious root formation may be under the control of the vegetative phase change pathway [44]. The ability of miR156 to promote adventitious root formation in Arabidopsis requires the auxin-induced expression of the *rtcs1/rtcl1* paralogs [50]. Collectively, these data demonstrate both a conserved function for juvenile determinants in the development of adventitious roots and a direct role for these regulators in root development.

We have curated a list of all genes with known mutant phenotypes affecting nodal roots, which we term the Identified Annotated Modulators of Brace Roots (IAMBROOT; **Table 1** and **Supplemental Table 1**). Consistent with the roles for determinants of phase transition in adventitious root formation, 14 of the 20 IAMBROOT genes also alter the juvenile-to-adult transition. Overexpression of miR156b/c in the maize *corngrass1* (*Cg1*) mutant extends the juvenile phase and represses multiple SPLs, including *teosinte glume architecture1* (*tga1*) [43]. Transgenic knockdown of TGA1 also resulted in an increase in nodes with brace roots [51]. In addition, loss of the miR172 target *zmrap2.7*, an AP2-LIKE transcription factor also known as *vegetative to generative transition1* (*vgt1*), results in reduced brace root emergence [25,52]. This indicates that the miR156 and miR172 signaling modules regulate brace root emergence.

Other genes with effects on phase transition outside of the miR156 and miR172 signaling modules are also in IAMBROOT (**Table 1** and **Supplemental Table 1**). For example, overexpression of *constans constans-like timing of cab1* (*zmcct1*), a major photoperiod response regulator, delays phase change and promotes brace root formation [53]. Loss of *early phase change* (*epc*) hastens the transition from juvenile-to-adult phase and results in fewer nodes with brace roots [54,55]. The gibberellic acid (GA) biosynthetic mutants, such as *dwarf plant1* (*d1*), *d3*, *d5*, and *anther ear1* (*an1*), extend the juvenile phase and increase the number of nodes with brace roots [38]. This is consistent with the observation that the DELLA-domain transcription factors that transduce GA signals repress *spl* genes, which explains the extended juvenile phase in the absence of GA [56,57]. These results highlight the intricate relationship of phase change, flowering time, and plant hormones in the timing and regulation of brace root development.

In addition to individual gene analyses, nodal root traits have been mapped as quantitative trait loci (QTL) in maize (**Supplemental Table 2**). The overlap between nodal root QTL and the IAMBROOT genes identifies candidates with known function in brace root development for 34 of the 161 QTL (**Table 2; Supplemental Table 3**). QTL affecting brace roots map to *rt1* [58] and *rtcl1* [59], as well as several genes known to regulate phase change, plant hormones, and flowering time. Brace root QTL overlapped with *zmcct1* and *zmrap2.7* [58–60], alleles of which comprise flowering time QTL in maize [61,62]. The over-representation of flowering time and phase change regulators in IAMBROOT suggests that additional candidates for brace root QTL may be found among known flowering time regulators. Therefore, we expanded upon IAMBROOT to include proposed flowering time regulators in maize from recent findings (**Supplemental Table 4**) and compared that list to the brace root QTL (**Supplemental Table 2**). This overlap identified 26 QTL (**Supplemental Table 5**) that includes *indeterminate1* (*id1*), *phosphatidylethanolamine-binding protein8/zeacentroradialis8* (*pebp8/zcn8*), *delayed flowering1* (*dlf1*), and *elongated mesocotyl1* (*elm1*), which encode *bona fide* flowering time QTL in maize [11,62] but have not been investigated for effects on brace root development. Direct investigation of these genes is required to determine if they are responsible for the associated QTL.

The mechanistic relationship between root developmental control and shoot meristem phase transitions may be fundamental to resource allocation in angiosperms [63]. Support for this hypothesis comes from a common garden experiment using 20 divergent dry-grassland species that found that flowering time predicted the allocation of root to shoot biomass better than phylogenetic relationships [64]. The fundamental association between the roles of roots in anchorage, water uptake, nutrient use, and now flowering time confounds the genetic analysis of brace root functions and the ability to optimize them for crop adaptation. Some experiments have resulted in independent effects on brace roots and flowering time [13,14,55], and molecular identification of these independent control mechanisms is required to test the functions of brace roots while controlling for variation in flowering time. Similar constraints on hypothesis testing are presented by the overlap between the control of juvenile-to-adult transition, adventitious root formation, lateral root emergence, and flowering time control. The shared mechanisms can result in pleiotropic effects of alleles on both flowering time and root development. Future experiments investigating the effects these genetic factors on both processes are critical for progress towards the rational design of climate and stress resilient crops.

## Aerial Nodal Root Function

Climate and stress resilient crops must have root systems that are optimized for plant anchorage, water uptake, and nutrient use in changing environments without compromising other physiological processes. Brace roots have been named for their importance in plant anchorage, which contributes to lodging resistance. Lodging, defined by the vertical displacement of stalks, can be a result of stem (stalk lodging) or root (root lodging) failure [65–68]. Lodging can be exacerbated by various factors, including groundwater accumulation, wind, drought, soil type, topographical variation, management practices, and plant growth stage [66,69,70]. Efforts to understand root lodging resistance in maize have identified correlations between the number of nodes containing brace roots that enter the soil and anchorage [11–13] and this function has now been demonstrated directly [14].

Beyond anchorage, brace roots are proposed to be vital for nutrient and water acquisition [9,71]. However, this assertion is based primarily on the anatomical presence of 36 or more metaxylem elements in brace roots compared to 6 metaxylem elements in the primary root and 15-18 metaxylem elements in crown roots [72]. Direct measures of water and nutrient uptake in brace roots as compared to crown roots have been limited. One study reports the hydraulic conductance of different maize root types using a pressure probe technique and includes both crown and brace roots [73]. In this study, the total hydraulic conductivity is the same for all root types, although the relative contribution of radial (across the root) to axial (up the root) conductance varied [73]. The lack of direct water and nutrient uptake measurements is likely due to the size and complexity of the root system when brace roots enter the soil. We performed a meta-analysis of the expression of nitrate transporters, ammonium transporters, and aquaporins in crown roots, brace roots in the soil, and brace roots that remain aerial from the Maize Developmental Atlas [74]. The expression of transporters varied between all three root types, but not in any one direction that would suggest increased nutrient or water uptake in any single root type (**Supplemental Figure 1**).

Although aerial brace roots may seem like an unlikely site for nutrient and water uptake since they do not enter the soil, a recent study showed that aerial brace roots may play an important role in nutrient acquisition [75]. This study showed that, in a maize landrace, diazotrophic (nitrogen-fixing) bacteria associate with the mucilage produced from aerial brace roots. The enrichment of nitrogen in this mucilage suggested that these microbial communities could contribute to the nitrogen demand of the plant [75]. Thus, aerial brace roots may also constitute an important site of nutrient and water acquisition. Additional research is necessary to fully define the relative roles of crown roots, brace roots in the soil, and aerial brace roots in nutrient and water acquisition.

## Prevalence of Brace Roots in *Poaceae*

The importance of brace roots for plant anchorage, water uptake, and nutrient use may be conserved in other genera. While, aerial nodal roots are present in several genera of *Poaceae* (the family that includes maize), there is limited research regarding their development and function.

As highlighted above, the mechanisms of nodal root induction and initiation are likely conserved, however emergence and growth differ between subterranean and aerial roots. The term "brace root" is narrowly restricted to roots that emerge above the soil and proposed function to anchor the plant. Using this definition, brace roots are present only in the *Andropogoneae* and *Paniceae* tribes (**Figure 2**). Here we present the limited research that has been reported in other genera. Future comparative studies between genera will be critical to define the shared and divergent development and function of aerial nodal roots in *Poaceae*.

Within the *Andropogoneae* tribe, the closest maize relative, *Sorghum*, shared a common ancestor 12-16 million years ago (MYA) [76,77]. Although there has been scant research on sorghum brace root development, brace root formation varies in sorghum (**Supplemental Table 2**)[78,79]. For example, the parents of a sorghum recombinant inbred line (RIL) population were shown to have a dramatically different number of nodes containing brace roots - Sansui (a Chinese landrace) has 6-8 nodes containing brace roots, whereas Jiliang 2 (an elite cultivar) only contains brace roots on the node closest to the soil [78]. Two QTL located on chromosome 6 and 7 explained 7.6% and 52.5% of the phenotypic variation, respectively [78]. A similar candidate gene search for sorghum brace root QTL indicates that the locus on chromosome 6 overlaps with the flowering time regulator *maturity1* [80]. Therefore, the shared control of brace roots and flowering time (**Supplemental Table 5**) may not be unique to maize.

The importance of understanding the mechanisms of nodal root development is further supported by sugarcane, which diverged from sorghum 6-9 MYA and is more closely related to sorghum than maize [81–83]. Sugarcane is primarily propagated vegetatively, which occurs when a piece of stem, termed a sett, is planted in the soil and the root system is derived from nodes only. Sugarcane shows remarkable plasticity in nodal root phenotypes, with roots that emerge from the sett showing fine, highly branched architectures, whereas the roots that emerge from the new shoot are thick and reminiscent of the nodal roots in maize and sorghum [84].

*Setaria*, commonly referred to as foxtail millet, is part of the *Paniceae* tribe, which is sister to the *Andropogoneae*. Like sorghum, maize, and sugarcane, *Setaria* also develops aerial nodal roots (**Supplemental Figure 2**). One recent study on crown root development in *Setaria viridis*, the wild relative of the domesticated *Setaria italica*, showed remarkable developmental plasticity in response to water availability [85]. Specifically, plants showed an increase in the number of crown root primordia but a decrease in crown root emergence in response to water deficit [85]. This response was conserved in the domesticated *Setaria italica*, the domesticated maize, and the wild relatives of maize, *Zea mays ssp. mexicana* and *ssp. parviglumis*. *Setaria* is an emerging model organism due to its small stature, diploid genome, and increasingly available molecular genetic resources [86–88]. Future studies in *Setaria* could rapidly advance our understanding of brace root development and function and provide a foundation for comparative studies within the *Poaceae*.

## Conclusion

Aerial nodal roots have been identified in several genera of *Poaceae*. In maize, these roots are called brace roots based on their critical function in plant anchorage, but their function has been poorly defined in other genera. The developmental mechanisms defined from current studies in maize provide a critical foundation for future comparative studies. Separating the regulation of brace root traits, phase change, and flowering time is essential to optimize brace roots without compromising other physiological processes.

The future of breeding resilient crops must consider dynamic responses of plant development and function to environmental stressors. Given its central function in water uptake, nutrient use, and anchorage the root system is an obvious target. Despite recent work highlighting a role in anchorage, the aerial portion of the root system has been poorly studied. Thus, it is of critical importance to explore aerial root development and function in greater detail in maize and other members of *Poaceae*.

## Acknowledgements

We thank Lindsay Erndwein of Illustrations by LindZeaMays for illustrating the figures of this article. We thank Dr. Andrew Doust for providing images of *Setaria* brace roots (**Supplemental Figure 2**). We also thank two anonymous reviewers and the associate editor for their insightful comments on this manuscript.

## Literature Cited

1. Lynch J: **Root architecture and plant productivity**. *Plant Physiol* 1995, **109**:7–13.

2. Aiken RM, Smucker AJM: **Root system regulation of whole plant growth**. *Annual Review of Phytopathology* 1996, **34**:325–346.

3. Hodge A, Berta G, Doussan C, Merchan F, Crespi M: **Plant root growth, architecture and function**. *Plant Soil* 2009, **321**:153–187.

4. Hochholdinger F, Park WJ, Sauer M, Woll K: **From weeds to crops: genetic analysis of root development in cereals**. *Trends in Plant Science* 2004, **9**:42–48.

5. Hochholdinger F, Woll K, Sauer M, Dembinsky D: **Genetic dissection of root formation in maize (*Zea mays*) reveals root-type specific developmental programmes**. *Annals of Botany* 2004, **93**:359–368.

6. Hochholdinger F: **The maize root system: morphology, anatomy, and genetics**. In *Handbook of maize: Its biology*. Edited by Bennetzen JL, Hake SC. Springer New York; 2009:145–160.

7. Feldman L: **The maize root**. In *The maize handbook*. . Springer; 1994:29–37.

8. Blizard S, Sparks EE: **Maize nodal roots**. In *Annual Plant Reviews online*. . American Cancer Society; 2020:281–304.

9. Hoppe DC, McCully ME, Wenzel CL: **The nodal roots of *Zea*: their development in relation to structural features of the stem**. *Canadian Journal of Botany* 1986, **64**:2524–2537.

10. Zhang Z, Zhang X, Lin Z, Wang J, Xu M, Lai J, Yu J, Lin Z: **The genetic architecture of nodal root number in maize**. *The Plant Journal* 2018, **93**:1032–1044.

11. Liu S, Song F, Liu F, Zhu X, Xu H: **Effect of planting density on root lodging resistance and its relationship to nodal root growth characteristics in maize (*Zea mays L.*)**. *Journal of Agricultural Science* 2012, **4**:182.

12. Sharma S, Carena MJ: **BRACE: A method for high throughput maize phenotyping of root traits for short-season drought tolerance**. *Crop Science* 2016, **56**:2996–3004.

13. Shi J, Drummond BJ, Habben JE, Brugire N, Weers BP, Hakimi SM, Lafitte HR, Schussler JR, Mo H, Beatty M, et al.: **Ectopic expression of ARGOS8 reveals a role for ethylene in root-lodging resistance in maize**. *The Plant Journal* 2019, **97**:378–390.

14. Reneau JW, Khangura RS, Stager A, Erndwein L, Weldekidan T, Cook DD, Dilkes BP, Sparks EE: **Maize brace roots provide stalk anchorage**. *Plant Direct* 2020, **4**:e00284.

**Tables**

**Table 1. Identified Annotated Modulators of Brace Roots in maize (IAMBROOT). Maize genes with mutant phenotypes known to influence brace roots.**

| Locus name | Gene Model (B73v4) | Root Phenotypes | Reference |
|---|---|---|---|
| *rootless1 (rt1)* | GRMZM2G163975 | Reduced number of nodal roots | Jenkins 1930 |
| *Teopod1 (Tp1)* | | Increase in number of brace root whorls | Poethig 1988 |
| *Teopod2 (Tp2)* | | Increase in number of brace root whorls | Poethig 1988 |
| *Teopod3 (Tp3)* | | Increase in number of brace root whorls | Poethig 1988 |
| *Hairy sheath frayed1 (Hsf1)* | GRMZM2G151223 | Increase in number of brace root whorls | Moose and Sisco 1994 |
| *dwarf plant1 (d1)* | GRMZM2G036340 | Increase in number of brace root whorls | Evans et al., 1995 |
| *dwarf plant3 (d3)* | GRMZM2G093195 | Increase in number of brace root whorls | Evans et al., 1995 |
| *dwarf plant5 (d5)* | GRMZM2G093603 | Increase in number of brace root whorls | Evans et al., 1995 |
| *anther ear1 (an1)* | GRMZM2G081554 | Increase in number of brace root whorls | Evans et al., 1995 |
| *lateral rootless1 (lrt1)* | | Shorter nodal roots at the seedling stage | Hochholdinger and Feix 1998 |
| *early phase change (epc)* | | Reduced number of nodal roots | Vega et al., 2002 |
| *Corngrass1 (Cg1)* | GRMZM5G838324/zmaMIR156 | Increase in number of brace root whorls | Chuck et al., 2007 |
| *rootless concerning crown and seminal roots1 (rtcs1)* | GRMZM2G092542 | No brace roots | Taramino et al., 2007 |
| *teosinte glume architecture1 (tga1) and neighbor of tga1 (not1)* | GRMZM2G101511/tga1; AC233751.1_FG002/not1 | Increase in number of brace root whorls | Wang et al., 2015 |
| *rtcs-like1 (rtcl1)* | AC149818.2_FG009 | Reduced elongation of shoot-borne crown roots | Xu et al., 2015 |
| *big embryo1 (bige1)* | GRMZM2G148937 | Increase in number of brace root whorls | Suzuki et al., 2015 |
| *zmcct1* | GRMZM2G381691 | Increase in number of brace root whorls | Stephenson et al., 2019 |
| *zmrap2.7* | GRMZM2G700665 | Reduced number of nodal roots | Li et al., 2019 |
| *argos8* | GRMZM2G354338 | Reduced number of nodal roots | Shi et al., 2019 |
| *vivaparous8 (vp8)* | GRMZM2G010353 | Increase in number of brace root whorls | Evans and Poethig 1997 |

**Table 2. Summary of the overlap between IAMBROOT candidate gene list and nodal root traits mapped in QTL analysis.** Candidate genes from the IAMBROOT curated list were overlaid with known nodal root QTL. QTL were compiled from Gu et al., 2016; Ku et al., 2012; and Zhang et al., 2018. A full list can be found in Supplemental Table 3.

| Annotation | Candidate Gene Model (B73 RefGenV4) | Root Traits | No. of QTL | IAMBROOT References |
|---|---|---|---|---|
| *an1* | Zm00001d032961 | BRN | 2 | Evans et al., 1995 |
| | | RNPL | 1 | |
| *argos8* | Zm00001d038075 | BRWN | 1 | Shi et al., 2019 |
| | | TNWN | 1 | |
| *cct1* | Zm00001d024909 | BRWN | 2 | Minow et al., 2018 |
| | | EBRWN | 1 | |
| | | ARNPL | 1 | |
| | | CRN | 1 | |
| | | CRNPL | 1 | |
| | | CRWN | 1 | |
| | | RNPL | 1 | |
| | | TNRN | 1 | |
| | | TNWN | 1 | |
| *d1* | Zm00001d039634 | BRWN | 1 | Evans et al., 1995 |
| | | TNWN | 1 | |
| *d5* | Zm00001d002349 | TNWN | 1 | Evans et al., 1995 |
| *rap2* | Zm00001d010987 | BRWN | 1 | Minow et al., 2018 |
| | | BRN | 2 | |
| | | CRN | 2 | |
| | | CRNPL | 1 | |
| | | TNWN | 1 | |
| *rt1* | Zm00001d040186 | BRN | 1 | Jenkins 1930 |
| | | BRWN | 1 | |
| | | TNRN | 1 | |
| | | TNWN | 1 | |
| *rtcl1* | Zm00001d048401 | BRWN | 1 | Xu et al., 2015 |
| | | EBRWN | 1 | |
| *rtcl1* | Zm00001d048401 | BRN | 1 | Xu et al., 2015 |

| | | BRN | 1 | |
| --- | --- | --- | --- | --- |
| *vp8* | Zm00001d034383 | TNRN | 1 | Evans and Poethig, 1997 |
| | | TNWN | 1 | |

ARNPL=Aerial Root Number Per Layer; BRN=Brace Root Number; BRWN=Brace Root Whorl Number; CRN=Crown Root Number; CRNPL=Crown Root Number Per Layer; CRWN=Crown Root Whorl Number; EBRWN=Effective Brace Root Whorl Number; RNPL=Nodal Root Number Per Layer; TNRN=Total Nodal Root Number; TNWN=Total Nodal Whorl Number

**Figures**

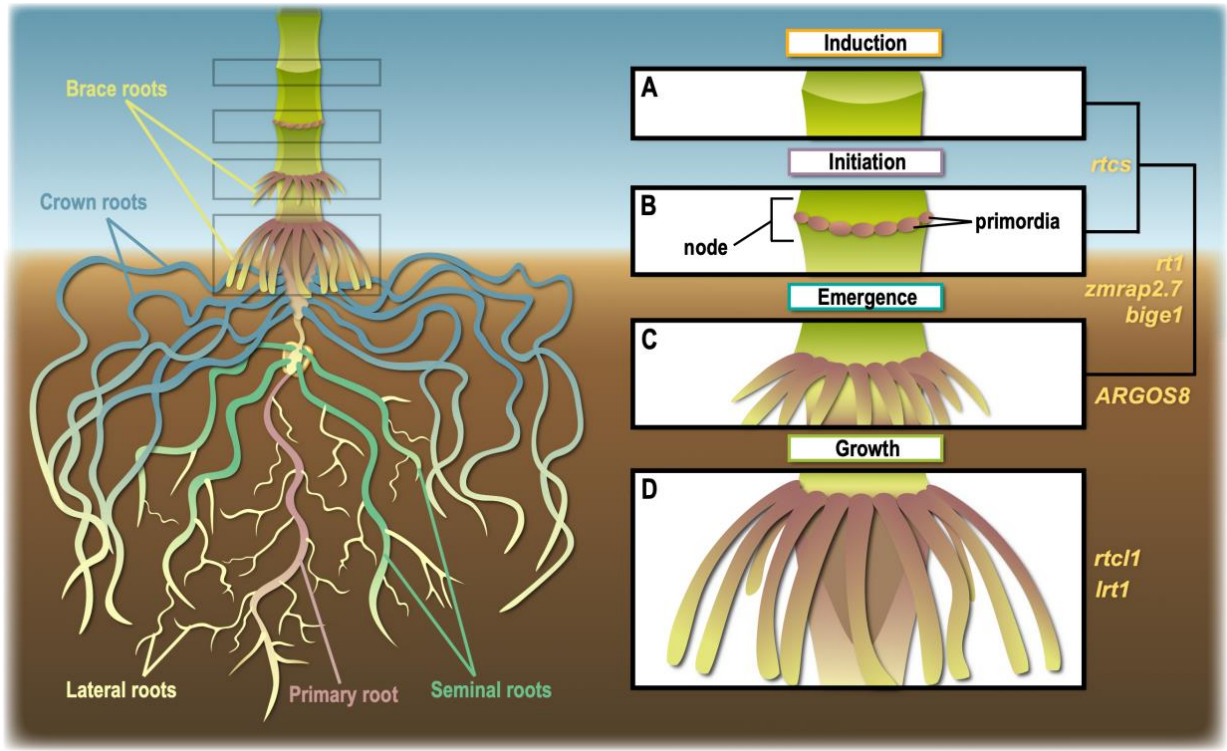

**Figure 1. Molecular development of maize nodal roots.** The maize root system consists of embryonic primary (red) and seminal (green) roots. The post-embryonic roots include the lateral (white), nodal crown (blue), and nodal brace (yellow) roots. Maize nodal roots develop from stems in four sequential stages as illustrated (A-D) for brace roots: induction, initiation, emergence, and growth. (A) The induction stage requires one or more cell types to become founder cells with the competency to divide; (B) once root primordia are established in stem nodes, plants enter the initiation stage; (C) roots then emerge from stems during the emergence stage; (D) root growth is the last stage of development. Genes that affect nodal root development are listed on the right. If the impacted stage of development has been determined, the genes are listed within the respective stages. If the specific developmental stage has not been determined, the genes are listed in brackets spanning the predicted stages.

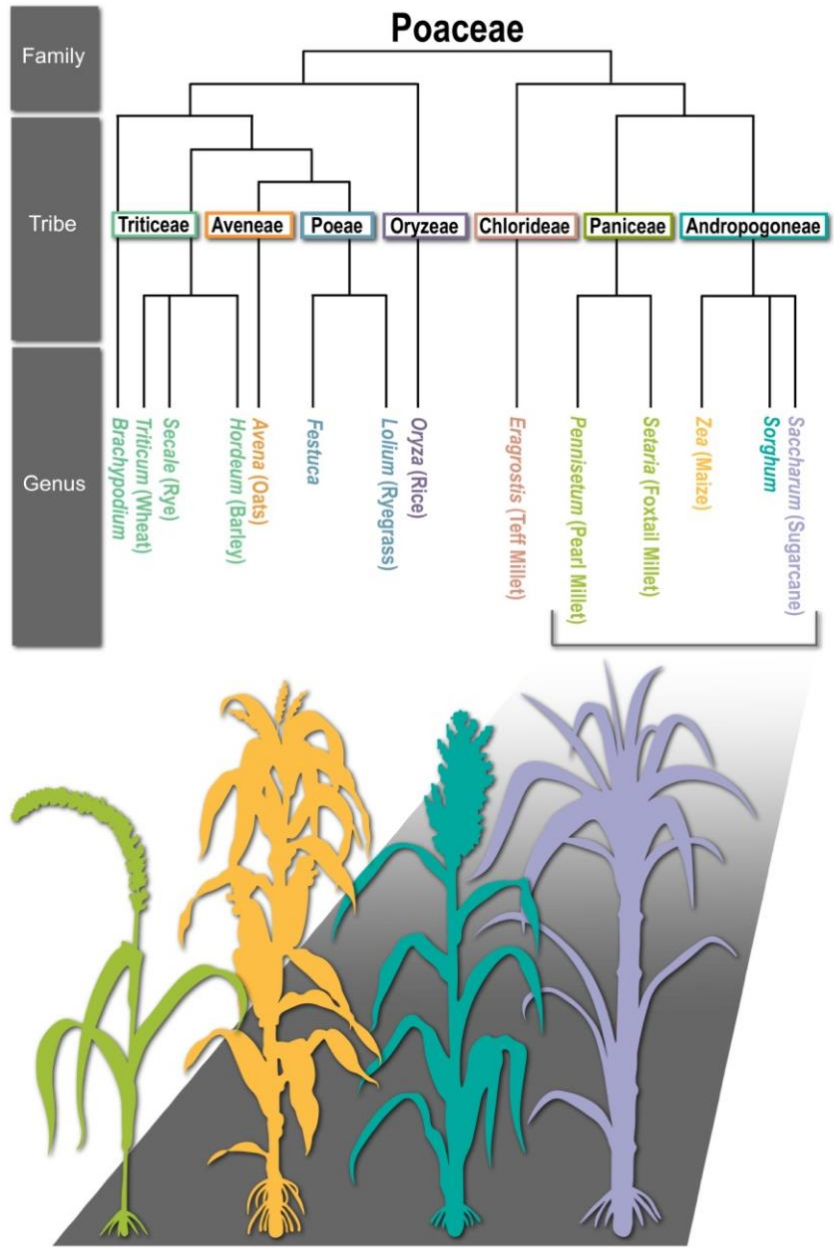

**Figure 2. Phylogenetic distribution of members in the *Poaceae* family.** Aerial brace roots have been observed within sugarcane (purple), sorghum (teal), maize (yellow), and *Setaria* (green), all of which are members of the *Andropogoneae* and *Paniceae* tribes. However, with the exception of maize, the development and function of these roots have been poorly defined.

## Supplementary Tables

**Supplemental Table 1.** Identified Annotated Modulators of Brace Roots (IAMBROOT)

**Supplemental Table 2.** Summary of all the nodal root QTL detected in various maize and sorghum studies

**Supplemental Table 3.** Summary of the overlap between the 161 QTLs reported in maize for nodal root trait variation and IAMBROOT gene list

**Supplemental Table 4.** Putative candidate gene list of maize genes that influence flowering time

**Supplemental Table 5.** Summary of the overlap between the 161 QTLs reported in maize for nodal root trait variation and bona fide flowering QTL in maize

# Supplementary Figures

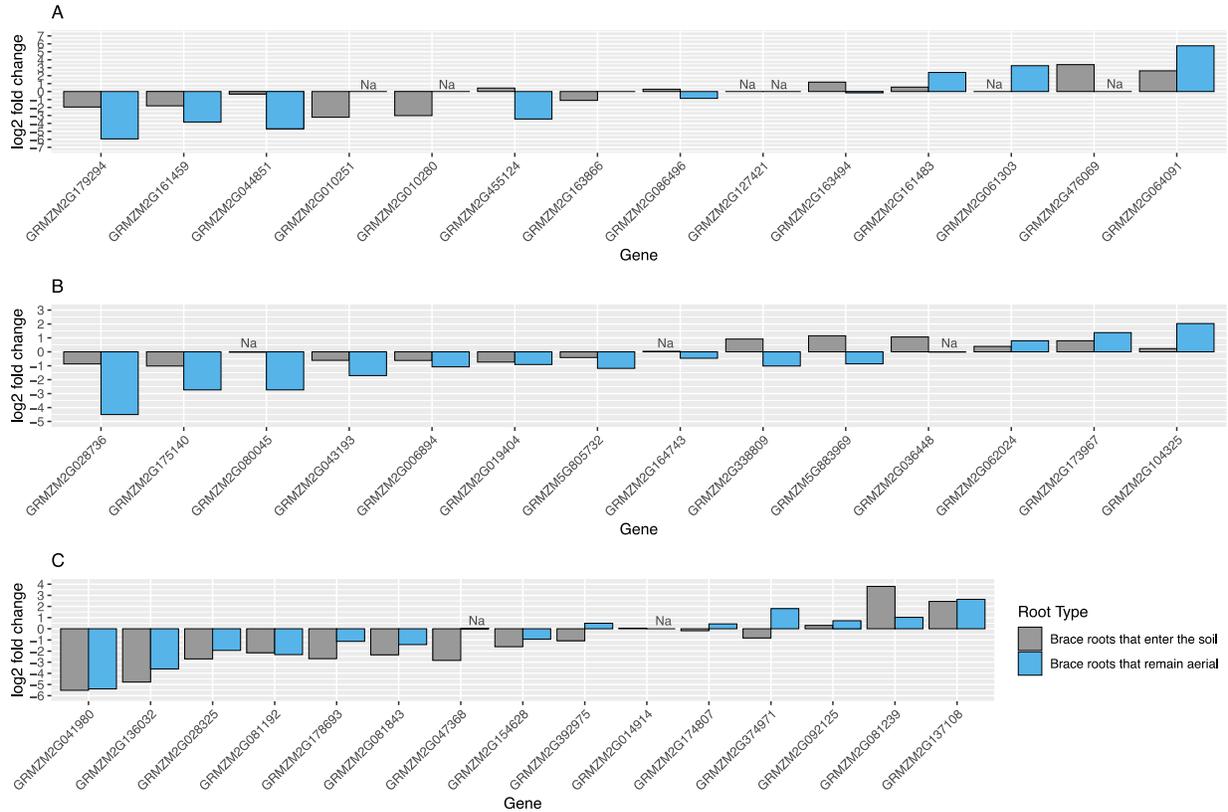

**Supplemental Figure 1. Transcriptome meta-analysis of nutrient and water transporters in maize nodal roots.** Data were mined from the Maize Developmental Atlas, considering crown roots [Crown_Roots_Node4_V7], brace roots that enter the soil [Crown_Roots_Node5_V13], and brace roots that remain aerial [Brace_Roots_Node6_V13] [74]. Although the brace that enter the soil were designated as crown roots in [73], these roots would be classified as brace roots by our definition – developing from nodes above the soil. The expression of transporters (counts per million, CPM) in brace roots was normalized by the expression in crown roots. Normalized expression values for brace roots that enter the soil and brace roots that remain aerial were converted to log2 fold change. (A) Normalized expression data for nitrate transporters (B) Normalized expression data for ammonia transporters (C) Normalized expression data for aquaporins. There is variable expression of transporters based on nodal root type. NA indicates expression was not detected in that root type.

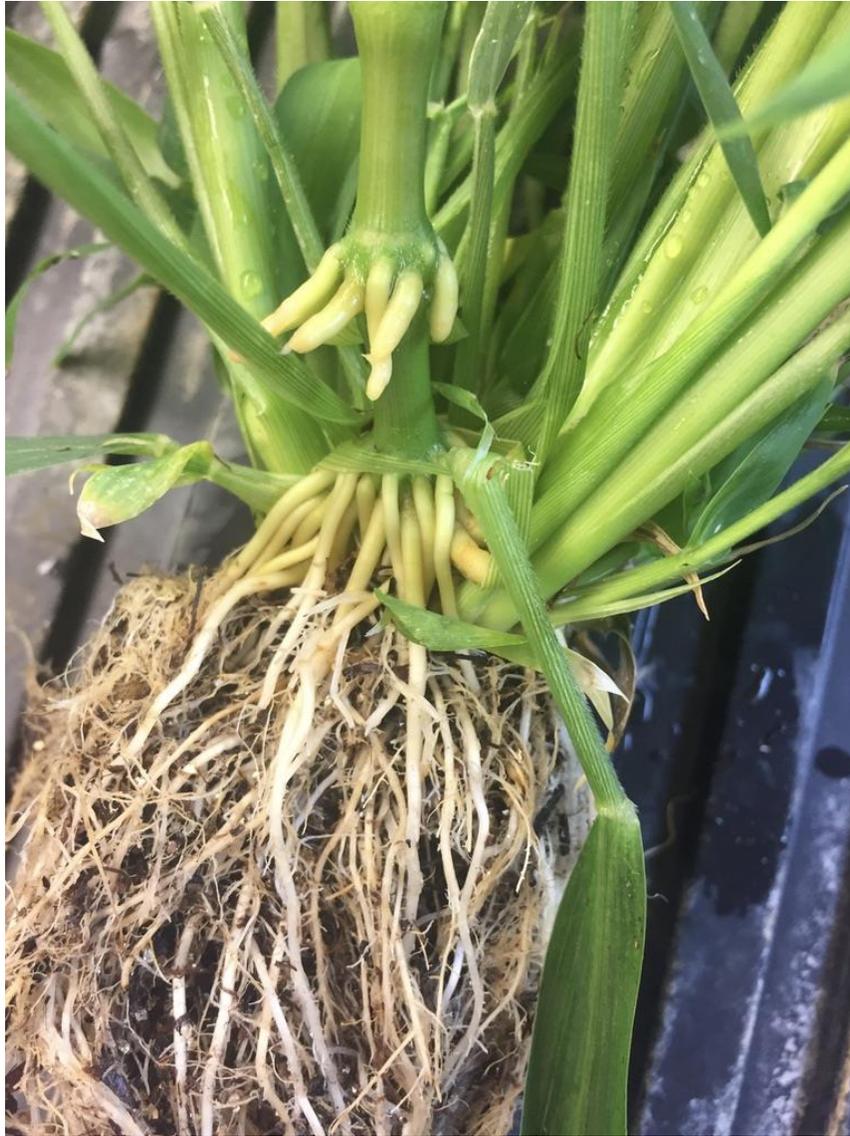

**Supplemental Figure 2. Brace root in *Setaria*.** Although not described in the literature, *Setaria* has distinct brace roots. Photograph courtesy of Dr. Andrew Doust (Oklahoma State University).